\documentclass[11pt]{article}

\makeatletter

\def\A{\mathsf{A}}
\def\B{\mathsf{B}}
\def\F{\mathsf{F}}
\def\K{\mathsf{K}}
\def\J{\mathsf{J}}
\def\SO{\mathsf{SO}}
\def\so{\mathsf{so}}
\def\g{\mathsf{g}}

\def\h{\mathsf{h}}

\def\p{\mathsf{p}}

\def\l{\mathsf{l}}

\def\f{\frac}

\def\extd{\mathrm{d}}

\def\tr{\mbox{Tr}}
\def\what{\widehat}

\def\be{\begin{equation}}
\def\ee{\end{equation}}
\def\bes{\begin{eqnarray}}
\def\ees{\end{eqnarray}}
\newcommand{\bea}{\begin{eqnarray}}
\newcommand{\eea}{\end{eqnarray}}
\@addtoreset{equation}{section}

\makeatother

\begin{document}

\title{%
Particles as Wilson lines of gravitational field}
\author{%
L.\ Freidel\thanks{ Perimeter Institute for Theoretical Physics, Waterloo, Canada;
Laboratoire de Physique, \'Ecole Normale
Sup{\'e}rieure de Lyon, Lyon , France; {\tt lfreidel@perimeterinstitute.ca}},%
$\;$ J.\ Kowalski--Glikman\thanks{Institute for Theoretical Physics,
University of Wroclaw,   Poland;  {\tt
jurekk@ift.uni.wroc.pl}},%
 $\;$ and A.\ Starodubtsev\thanks{Institute for Theoretical
 Physics,
University of Utrecht, The Netherlands;  {\tt
A.Starodubtsev@phys.uu.nl}}%
} \maketitle

\begin{abstract}
Since the work of Mac-Dowell-Mansouri it is well known that gravity
can  be written as a gauge theory for the de Sitter group. In this
paper we consider the coupling of this theory to the simplest gauge
invariant observables that is, Wilson lines. The dynamics of these
Wilson lines is shown to reproduce exactly the dynamics of
relativistic particles coupled to gravity, the gauge charges carried
by Wilson lines being the mass and spin of the particles. Insertion
of Wilson lines breaks in a controlled manner the diffeomorphism
symmetry of the theory and the gauge degree of freedom are
transmuted to particles degree of freedom.
\end{abstract}

\clearpage

\section{Introduction}

``Geometry tells matter how to move; matter tells geometry how to
curve'' --  this simple sentence encompasses the main physical
message of general theory of relativity. Yet, as any great idea, it
contains the seed of problems: there is an explicit dichotomy in
this statement -- the division of the physical world into two
entities, matter and geometry.

There is an old dream, preceding the idea of unification of all
interactions,  to unify geometry and matter, usually in the guise
of geometrization of matter (though string theory seems to take
the completely opposite stance, attempting to ``matterize''
gravity.) If gravitational field {\em is} geometry is it that not
reasonable to expect that all other physical objects could be
described in terms of geometry of some sort as well? In this paper
we aim to give an affirmative answer to this question. To be more
precise we show that point particles, with momentum and spin,  can
be described as Wilson lines of an appropriate connection.

The particle can be described by the momentum and spin it carries.
These are charges of Poincar\'{e} group, which is also the gauge
group of gravity with vanishing cosmological constant. However
recent  developments indicate that in order to construct a
reasonable theory of (quantum) gravity one presumably should include
an infrared regulator in the form of cosmological constant. Thus the
minimal geometrical model must be based on the gauge group being de
Sitter group, $\SO(4,1)$ in four dimensions. The vanishing (or small
-- as it is the case in our universe) cosmological constant setting
can be obtained then by taking an appropriate limit of such a
theory. Fortunately the charges of this group can be still
interpreted as momentum and spin, so coupling of gravity understood
as a gauge theory of $\SO(4,1)$ can be naturally coupled to matter.
This also raised the possibility that matter could be understood in
terms of some specific configurations of gravitational field. In
this paper we show how this construction can be realized explicitly.

In the next section, following \cite{FreidelA} we recall how gravity
can  be constructed as a constrained topological field theory. In
section 3, using the formalism of Balachandran et.\ al.\
\cite{BalLNP}  we shortly review the way the point particles can be
coupled to external gauge field. We show that in the case of the
$\SO(4,1)$ gauge group the resulting equations reduce to the
standard Mathisson--Papapetrou form, in the appropriate limits. In
the following section we show that also the gravitational field
equations acquire the correct form, with energy momentum tensor, and
the source of torsion being point-like particles with appropriate
energy-momentum and spin. Section 5 explains how particles could be
understood in terms of Wilson lines. We conclude this paper with
some comments, and with an appendix containing some useful formulas.

\section{The action of gravity and topological field theory}

In this section, following \cite{FreidelA} we present for the
reader's convenience the action for gravity as an action of a
perturbed topological BF system. This formulation is an extension of
the original work of Mac Dowell and Mansouri \cite{mm} who expressed
gravity as a $\SO(4,1)$ gauge theory. We focus on $\SO(4,1)$  since
it concerns Lorentzian gravity with a positive cosmological constant
which is clearly the most physically interesting case. Everything we
say in this paper can be readily translated to the case of negative
cosmological constant or Euclidean gravity. Note however that we
definitely need a non zero cosmological constant for this formalism
to make sense.

The gravitational field is encoded it in a $\SO(4,1)$ (de Sitter)
connection \be \A_\mu= \left(\f{1}{2l}e_\mu{}^a\,{\gamma_a \gamma}
+\f{1}{4} \omega_\mu{}^{ab}{}\, \gamma_{ab}\right) \ee where the
index $a$ runs from $0$ to $3$. In this formula, gamma matrices
$\gamma_a \gamma/4$, $\gamma_{ab}/4$ form a representation of
generators $T^{IJ}$ ($I,J = 0,\ldots 4$, $\gamma = \gamma_4$) of the
Lie algebra $\so(4,1)$. $e_{\mu}{}^{a}$ is the frame field from
which the metric is constructed and $\omega_{\mu}{}^{ab}$ is the
spin connection. When written in terms of the $\so(4,1)$ generators
the connection reads $\A= A^{IJ}{T_{IJ}}$. Connection has a mass
dimension 1 whereas the frame field  is dimensionless; this is the
reason why a length scale $l$ appears in the expression of the
components of $\A$ representing the frame field. As we will see in
order to recover the usual gravitational dynamics this length scale
has to be the cosmological length $l$ related to the cosmological
constant by \be \frac{1}{l^2}= \frac{\Lambda}{3} \ee

To formulate the theory  we also need a two form valued in the Lie
algebra $\so(4,1)$ denoted by $\B = B_{\mu \nu}{}^{IJ}{T_{IJ}}\,
\extd x^\mu\wedge \extd x^\nu$. In terms of this two form and
connection $\A$ the action takes the form
\begin{equation}\label{2.1}
    -\frac12 S = \int\, \mbox{Tr } \left(\B \wedge \F(\A)
    -\frac{i\alpha}{2}\, \B \wedge \B\, \gamma - \frac{\beta}{2}\, \B \wedge \B\right)
\end{equation}
which can be rewritten in components as
\begin{equation}
S=\int \left(B_{IJ}\wedge F^{IJ} -\frac{\alpha}{4}\, B_{IJ}\wedge
B_{KL}\epsilon^{IJKL4}-  \frac{\beta}{2}\, B^{IJ}\wedge
B_{IJ}\right) \label{2.1a}
\end{equation}
This action can be shown \cite{FreidelA} to be equivalent to the
action of General Relativity, for $\alpha\neq0$.

 In order to solve the equation of motion for $B$ it is convenient
to introduce the operators
\begin{equation}\label{e12}
    {\cal P}_{\pm}{}^{ij}{}_{ kl} =  \left[\pm\frac\alpha2\,\epsilon^{ij}{}_{kl}  +\beta\, \delta^{ij}_{kl}\right],
    \quad {\cal P}_{-}{}^{ij}{}_{ kl}{\cal P}_{+}{}^{kl}{}_{ mn}=({\alpha^2+\beta^2})\delta^{ij}_{mn}
\end{equation}

The field equations for $\B$ following from (\ref{2.1}) read
 \bea
 B^{ij} &=& \frac{1}{\alpha^2 +\beta^2} {\cal P}_{-}{}^{ij}{}_{ kl} F^{kl},\\
B^{4i} &=& \frac{1}{\beta } F^{4i}. \eea These equations are
algebraic, so we can substitute them back to (\ref{2.1}) to obtain
\be\label{action42} S=\int \left(
\frac{-\alpha}{4(\alpha^2+\beta^2)} F^{ij}\wedge
F^{kl}\epsilon_{ijkl} + \frac{\beta}{2(\alpha^2+\beta^2)}
F^{ij}\wedge F_{ij} + \frac{1}{\beta} F^{5i}\wedge F_{5i}\right) \ee
The curvatures of connection $\A$ are decomposed as follows
\begin{eqnarray}
F^{ij}(\A)=R^{ij}(\omega)- \frac{1}{\l^2}e^i \wedge e^j \nonumber
\\
F^{i4}(\A)=\frac{1}{\l}d_\omega e^i \label{fdecomp}.
\end{eqnarray}

Using (\ref{fdecomp}) and introducing the Nieh-Yan class $$C=
d_\omega e^i \wedge d_\omega e_i - R^{ij}\wedge e_i\wedge e_{j}$$
 we can rewrite this
action in terms of gravity variables
$$\label{topclass}
S=\widetilde{S}_P - \frac{\alpha}{4(\alpha^2+\beta^2)}\, \int
R^{ij}(\omega) \wedge R^{kl}(\omega)\epsilon_{ijkl}$$
\be\label{action43} +\frac{\beta}{2(\alpha^2+\beta^2)}\, \int
R^{ij}(\omega) \wedge R_{ij}(\omega) +\frac{1}{\beta}\, \int C. \ee
The last three terms are the integrals of   Euler,  Pontryagin, and
Nieh-Yan classes. These are integer valued topological invariants
with trivial local variation. The first term of action
(\ref{action43})
$$
\widetilde{S}_P = \frac{1}{2G}\int
 R^{ij}(\omega)\wedge e^k \wedge e^l \epsilon_{ijkl}$$ \be -
\frac{\Lambda}{12G}\int e^i \wedge e^j \wedge e^k \wedge e^l
\epsilon_{ijkl} +\frac{1}{G\gamma}\int R^{ij}(\omega)\wedge e_i
\wedge e_j
 \label{actionfinal}
\end{equation}
is the  action  of General Relativity with nonzero cosmological
constant and a nonzero, dimensionless Immirzi parameter
 $\gamma$.
The initial parameters $\alpha, \beta, \l$ are related to the
physical ones as follows \be\label{parameter}
\frac{1}{\l^2}=\frac{\Lambda}{3},\qquad \alpha =
\frac{G\Lambda}{3}\frac{1}{(1+\gamma^2)}, \qquad \beta =
\frac{G\Lambda}{3}\frac{\gamma}{(1+\gamma^2)}. \ee Even if the term
proportional to $\gamma^{-1}$ in (\ref{actionfinal}) is not
topological (its variation is non zero), it doesn't affect the
classical equation of motion when $\gamma^2 \neq -1$ unless the
theory is coupled to fermions \cite{fermions}. We recover the usual
metric gravity in the case $\gamma=0$, in this case the torsion is
forced to vanish. The other extreme $\gamma=\infty$ correspond to
Cartan-Weyl formulation of gravity \cite{FreidelA}. It is important
to note that in both cases we have $\beta=0$.

 We see therefore that for $\alpha\neq0$ the action  (\ref{2.1})
reproduces the action of General Relativity accompanied with a
number of topological terms. This action makes it possible also to
consider limits $\alpha\rightarrow0$, in which equations of motion
of General Relativity turn into those of topological field theory.
Let us first consider $\alpha=0$, $\beta=0$ limit  of (\ref{2.1})
which, as it follows from (\ref{parameter}), corresponds to the
limit $G\to 0$. In this case we have to do with a pure  $BF$ theory
described by the action
\begin{equation}\label{2.2}
      -\frac12 S_0 = \int\, \mbox{Tr } \B \wedge \F(\A),
\end{equation}
whose equations of motion simply state that geometry is de Sitter
flat \be F(A)=0 \ee.  The topological theory (\ref{2.2}) is
invariant under two gauge symmetries, the standard
\begin{equation}\label{2.3}
    \A \mapsto \g^{-1} \A \g + \g^{-1} d \g, \quad \B \mapsto \g^{-1} \B \g
\end{equation}
along with
\begin{equation}\label{2.4}
\A\mapsto \A ,\quad
 \B\mapsto \B + d_\A \Phi
\end{equation}
where $d_{\A}={d} + {[}\A,\cdot{]}$ is the covariant exterior
derivative and $\Phi$ being an $\so(4,1)$ valued one form.

Another possible limit is $\alpha=0$, $\beta\neq0$, which involves
large Immirzi parameter
\begin{equation}\label{2.2a}
      -\frac12 S_0 = \int\, \mbox{Tr } \left(\B \wedge \F(\A) - \frac{\beta}{2}\, \B \wedge \B\right)
\end{equation}
This theory is also invariant under (\ref{2.3}), while (\ref{2.4})
is replaced with
\begin{equation}\label{2.4a}
\A\mapsto \A +\beta \Phi,\quad
 \B\mapsto \B + d_\A \Phi + \frac{\beta}{2} {[}\Phi,\Phi{]},
\end{equation}
which implies that spacetime geometry is arbitrary in the bulk.

 In what follows we will consider mainly the most general case
$\alpha,\beta\neq 0$, corresponding to the full dynamical gravity
theory (with torsion and Immirzi parameter.) In some instances we
will also discuss particular topological limits of this theory.

\section{Particle action and equations of motion}

In the formulation of the previous section gravity is formulated as
a $\SO(4,1)$ gauge theory, of which only the $\SO(3,1)$ part is
unbroken by the  gravitational term proportional to $\alpha$. As
shown in \cite{LeeA,FreidelA} in the case of pure gravity, the
formalism can be extended to be $\SO(4,1)$ gauge invariant with
spontaneous symmetry breaking down to $\SO(3,1)$ Lorentz gauge
invariance. The first goal of this paper is to show that matter can
arise in the most natural way in this formalism by introducing the
simplest possible term breaking the gauge symmetry of the theory in
a localized way. The gauge degrees of freedom are then promoted to
dynamical degree of freedom, and as we will show, reproduce the
dynamics of a relativistic particle coupled to gravity. This
realizes explicitly in four dimension the idea that matter
(relativistic particles) can arise as a charged (under $\SO(4,1)$)
topological gravitational defect. This strategy, well known in three
dimensions,  gives a new perspective where matter and gravity are
geometrically unified \cite{desousa} and was the key ingredient in
the recent construction of the effective  action of matter fields
coupled to quantum three dimensional gravity \cite{FreidelL}.

An equivalent way to present  the inclusion of matter in our context
is to  realize that the only natural way to couple a gauge field to
localized excitation is by insertions of Wilson lines. Remarkably,
the dynamics of these Wilson lines is the one of a relativistic
particles. The formalism that allows to reach this conclusion was
first developed by Balanchandran et.\ al.\ \cite{BalLNP}.
 In this section we consider spinning particle moving
in an external gravitational field, the full description of the
particle(s) --  gravity system will be described in the next
section, while we return to Wilson lines in Section~5.

The simplest possible localized gauge breaking  coupling to the
gravitational  field\footnote{We restrict in this paper to the gravitational coupling which has a clear physical interpretation.
The coupling of string like sources to the $B$ field in $BF$ theory
as been recently considered in \cite{AB}, but its physical interpretation in the full theory is far from clarified.}
$\A$ is obtained by choosing a worldline $P$
and a fixed element $K$ of the $\so(4,1)$ Lie algebra so as to have
\be S_{P}(\A) =-\int \extd \tau\, \tr \left(\K \A_\tau(\tau)\right),
\ee where $\tau$ parameterizes the world line $z^\mu(\tau)$ and
$\A_\tau(\tau) \equiv \A_\mu(z(\tau))\, \dot{z}^\mu$.

 This action breaks  gauge
invariance and diffeomorphism symmetry\footnote{ This is true for diffeomorphism that modify the worldline location.
The action is still invariant under the residual symmetry consisting of diffeomorphisms that acts along the worldline,
i-e reparametrisation invariance.}
 In order to restore the symmetry at the
particle location one promotes the gauge degree of freedom to the
dynamical ones, which can be interpreted as Lorentz frame and
particle position. This is similar to what happens in three
dimension.

The 4-dimensional de Sitter group  acts by conjugation on its Lie
algebra,  the orbits of this action being labelled by two numbers
$(m,s)$ which are the mass and spin of the particle. For each orbit
we choose a fixed representative element of the 4-dimensional de
Sitter Lie algebra (for conventions see the Appendix)
\be\label{2a.2} \K\equiv m l {\gamma^0\gamma}/{2} + s
{\gamma^2\gamma^3}/{4} \ee with obvious generalization in the case
of massless particles.

The Lorentz Lie algebra $\so(3,1)$ is identified with the subalgebra
of $\so(4,1)$ generated by $\gamma_{ab}$. The Lagrangian of a single
particle propagating in a gravitational field is characterized by an
embedding of its worldline $z(\tau)$ and a function $\h(\tau)$
valued in the Lorentz subgroup $\h =
\exp(\alpha^{ab}{\gamma_{ab}}/{4})$. This function represents a
Lorentz transformation from the rest frame, in which the Poincar\'e
charges of the particle are described by the algebra element
(\ref{2a.2}) to an actual frame, in which the particle has momentum
$p$ and spin $s$ (see eq.\ (\ref{2a.7}) below.)

Let us denote by $\A^\h=\h^{-1} \A \h + \h^{-1} d \h$ the
corresponding gauge  transformation of $\A$. Then the lagrangian
takes the simple form \be\label{1.1} L(z,\h; \A) = -\tr \left(\K
\A^\h_\tau(\tau)\right)\quad S = \int\, d\tau\, L(z,\h; \A). \ee
 This lagrangian can be rewritten also in the following form
\begin{equation}
L(z,\h; \A)=-{\tr}(\J \A_\tau) + L_1(z,\h) \label{actionpt}
\end{equation}
where in the first term  $\J$ is given by \be\label{J} \J \equiv
\h\, {\K} \,\h^{-1}. \ee The components of $\J$ can be expressed in
terms of the particle's  momenta and spin \be\label{2a.7} \J=
\f{\l}{2}p_a\,{\gamma^a \gamma} + \f{1}{4}s_{ab}{}\, \gamma^{ab}.
\ee The first term in equation (\ref{actionpt}) describes the
covariant coupling between the particle and the $\A$ connection of
the (constrained) BF theory, while the second
\begin{equation}
L_1(z,\h) =  -\tr(\h^{-1}\dot \h \K), \label{actpartd}
\end{equation}
describes the dynamics of the particle.

This action is analogous to the spin part of the action
 for particle in three dimensional gravity\cite{desousa}. The
difference is that the gauge group is now $\SO(4,1)$ which has two
Casimir operators, mass and spin, and the information about the two
Casimir operators is encoded in the extrinsic source $\K$,
(\ref{2a.2}).

To put (\ref{actionpt}) into more conventional form let us rewrite
it explicitly distinguishing the rotation transformations generated
by $\gamma_{ab}$, $a,b=0,\ldots,3$ and 'translation' transformations
generated by $\gamma_a \gamma $. By introducing the scalars
$$J^{IJ}=-{\tr }(\J \gamma^{IJ}/2), \quad \J=J^{IJ}\gamma_{IJ}/4$$
recalling that $A^{a4}= \sqrt{\Lambda/3}e^{a}$ and  introducing
momentum $p^a=\sqrt{\Lambda/3} J^{a4}$, and spin $s^{ab} = J^{ab}$
we can rewrite (\ref{actionpt}) as
\begin{equation}
L(z,\h; \A)= \frac12(A_\tau^{IJ}J_{IJ}) + L_1(z,\h) = e_\tau^a p_a +
\frac{1}{2}\omega_\tau^{ab} s_{ab} + L_1(z,\h) \label{actionptd}
\end{equation}

Since $\J$ in the above equations is an element of the $\so(4,1)$
algebra, $J^{IJ}$ must satisfy the constraints
\begin{equation}\label{C1}
\frac{1}{2}J^{IJ} J_{IJ}=C_2 \label{c2}
\end{equation}
and
\begin{equation}\label{C2}
\frac{1}{16}J^{IJ}J^{KL}\epsilon_{IJKLM}\epsilon^{MABCD}J_{AB}J_{CD}=C_4,
\label{c4}
\end{equation}
 where $C_2$ and $C_4$ are eigenvalues  of quadratic and
quartic Casimir operators of $\so(4,1)$ algebra.

To see what is the physical meaning of $C_2$ and $C_4$ let us
rewrite the equations (\ref{c2}) and (\ref{c4}) using the notations
as in (\ref{actionptd}). Assuming that the cosmological constant
$\Lambda$ is small and considering the leading order in $\Lambda$ we
get
\begin{equation}
C_2=\frac12 J^{ab} J_{ab}+J^{4 a}J_{4 a}=\frac12s^{ab}
s_{ab}+\frac{3}{\Lambda}p^a p_a \approx \frac{3}{\Lambda}p^a p_a
\label{c2a}
\end{equation}
and
\begin{eqnarray}
C_4&=&(\frac14 J^{ab} J^{cd} \epsilon_{abcd})^2+J^{ab} J^{4 c} \epsilon_{abcd}\epsilon^{d efg} J_{4 e }J_{fg} \nonumber  \\
&=&(\frac14 s^{ab} s^{cd} \epsilon_{abcd})^2+\frac{3}{\Lambda}s^{ab}
p^{ c} \epsilon_{abcd}\epsilon^{defg} p_{e }s_{fg}\nonumber \\
&\approx& \frac{3}{\Lambda}s^{ab} p^{ c}
\epsilon_{abcd}\epsilon^{defg} p_{ e }s_{fg} \label{c4a},
\end{eqnarray}
which is proportional to the length of the Pauli-Lubanski vector.

From the way the particles are coupled to the connection in
(\ref{actionptd}) it is clear that $p_\mu \equiv e_\mu{}^a\, p_a$ in
the above equations is space-time momentum.  From (\ref{c2a}) one
can see that the Casimir $C_2$ gives rise to the mass of the
particle
\begin{equation}
m^2=\frac{\Lambda }{3}C_2.
\end{equation}
The last equation for the mass relates two well-known problems in
particle physics.  Since the representation theory is labelled by
integers, there is a natural unit and the most natural choice is to
take the representation with minimal $C_2$.  Under this assumption,
explaining why the cosmological constant is small would also help to
explain why masses of elementary particles are small, and vice
versa.

In the center of mass frame, $p_a= (m,0,0,0)$, the Casimir $C_4$ in
(\ref{c4a}) can be rewritten as
\begin{equation}
C_4=\frac{3m^2}{\Lambda}s^{ij}\epsilon_{ijk}\epsilon^{klm}s_{lm}=C_2
s^i s_i,
\end{equation}
where $i,j, \ldots =1,2,3$ are $\SO(3)$-indices and
$s^i=\epsilon^{ijk}s_{jk}$ is the  spin in the rest frame of the
particle. Thus we have the expression for the spin
\begin{equation}
s^2=\frac{C_4}{C_2}.
\end{equation}
\vspace{12pt}

Consider now the equations of motion that follow  from the
Lagrangian (\ref{1.1}). The variation over $\h$ gives (ignoring
total derivatives) \bes\label{var1}
\delta L &=& -\tr\left(\h^{-1}\delta\h ([\K,\A^\h])\right),\\
         &=& -\tr\left(\delta\h \h^{-1} (D_\tau \J)\right),
\ees where we introduced the $\SO(4,1)$ covariant derivative along
the world-line of the particle
\begin{equation}\label{1.2}
     {D}_{\tau} \equiv \frac{d}{d\tau} + {[}\A_\tau,\cdot{]},
\end{equation}
and $\J$ is defined by equation (\ref{J})

Since $\h$ is restricted to be in the Lorentz subgroup $\SO(3,1)$ of
$\SO(4,1)$
 equations (\ref{var1}) constrain only the spin part of $\J$
and give the spin precession equation \be\label{sprec} D_\tau J^{ab}
= \nabla_\tau s^{ab} + e_\tau{}^a p^b -e_\tau{}^b p^a =0, \ee with
$\nabla_\tau \equiv \frac{d}{d\tau} + {[}\omega_\tau,\cdot{]}$ the
Lorentz connection and $e_\tau{}^a  = e_\mu{}^a \dot{z}^\mu$. Note
that by construction the momenta and spin satisfy the orthogonality
condition \be\label{ps} s^{ab}p_b =0. \ee In what follows we will
also need the translational part of the current derivative
\be\label{transc} \frac{1}{l}D_\tau J_{a} = \nabla_\tau p_{a}
-\frac{1}{l^2} s_{ab}e_\tau{}^b \ee

The variation over $z$ gives \bes\label{var2}
\f{\delta L}{\delta z^\mu} &=& \f{d}{d\tau}\tr(\J \A_\mu) - \tr(\J\partial_\mu \A_\nu)\dot{z}^\nu,\nonumber\\
         &=& \tr\left( D_\tau \J \A_\mu \right) - \tr\left(\J F_{\mu \nu}(\A)\right)\dot{z}^{\nu}=0\label{var21}
\ees where \bes
&&F_{\mu \nu}(\A)\equiv \partial_{\mu} \A_{\nu} -\partial_{\nu}\A_\mu +{[}\A_\mu,\A_\nu{]},\nonumber\\
 & = & T_{\mu \nu}{}^a \gamma_a\gamma /2 + \left(R_{\mu \nu}{}^{ab}(\omega) -\frac{1}{l^2}( e_\mu{}^a e_\nu{}^b
 -e_\nu{}^a e_\mu{}^b)\right)\gamma_{ab}/4
 \ees Here
$$T^a = de^a +\omega^a{}_b\wedge e^b = T_{\mu\nu}{}^{a} dx^{\mu}\wedge dx^{\nu}/2$$ is the torsion,
while $$R^{ab}= d\omega +[\omega,\omega]/2=R_{\mu\nu}{}^{ab}
dx^{\mu}\wedge dx^{\nu}/2$$ is the Lorentz curvature.

If one uses  equation (\ref{sprec}),  equations (\ref{var21})
written in components reads \be (\nabla_\tau p_a )e_{\mu}{}^{a} =
\f{1}{2}s_{ab}\, R_{\mu\nu}{}^{ab}\, \dot{z}^{\nu} + p_a T_{\mu
\nu}{}^a\,\dot{z}^{\nu}. \ee This is  Mathisson--Papapetrou
\cite{MP} equations describing the dynamic of spinning particle in
the presence of torsion  in an arbitrary gravitational background.
When  torsion is zero we recover the usual Mathisson-Papapetrou
equation, when spin is also zero we recover the usual geodesic
equation.

 This equation can be written in the more usual form if one
introduces the affine connection $\Gamma_{\mu \nu}{}^\rho$, which is
related to the spin connection $\omega_\mu^{ab}$ by the identity $
\partial_{\mu} e_\nu{}^a + \omega_{\mu}{}^a{}_b e^b_\nu =
\Gamma_{\nu \mu}{}^\rho e_\rho{}^a$. It can be written in terms of
the Christofell symbol $\what{\Gamma}$ as $\Gamma_{\mu \nu\rho} =
\what{\Gamma}_{\mu \nu\rho} + T_{\rho \{\mu \nu\}} -\f{1}{2}T_{\mu
\nu \rho}$ where $T_{\mu\nu\rho}= T_{\mu \nu}{}^{a}e_{\rho a}$ is
the torsion tensor, and the Mathisson--Papapetrou equation reads \be
\nabla_\tau p_\mu =\f{1}{2}\,s_{ab}\,R_{\mu\nu}{}^{ab}\,
\dot{z}^{\nu}, \ee where $p_\mu\equiv p_a e_\mu{}^a$ and $\nabla_\mu
p_\nu \equiv
\partial_{\mu}p_\nu -\Gamma_{\mu \nu}{}^\rho\, p_\rho$.

We see therefore that the $\SO(3,1)$ gauge transformation and the
diffeomorphism symmetry is restored at the particle location while
the gauge parameters acquire physical meaning being the Lorentz
frame and particle position.\newline

 Above we restricted ourselves to $\h$ being valued in the
Lorentz subgroup $\SO(3,1)$ of the full de Sitter group
$\SO(4,1)$. We know however that if one further restrict ourselves
to the topological case where $\alpha =0$, the bulk action is also
invariant under de Sitter gauge transformations. In this case we
can have a de Sitter covariant formulation of particle dynamics
similar to that in 2+1 gravity, where $\h$ has to be taken an
element of $\SO(4,1)$.  Let us try therefore to take $\h \in
\SO(4,1)$, and see which additional equation will result. In doing
so we would get from (\ref{var1}) an additional equation
 \be\label{hproblem}
 \nabla_\tau p_a = \frac{1}{l^2}\, s_{ab}e_\tau{}^b.
 \ee
This equation is equivalent to eq.\ (\ref{var21}) provided that
the following identity is satisfied along the particle worldline
\be \frac{1}{l^2}s_{ab}\,
e_{\mu}{}^{a}\,e_\tau{}^b=\f{1}{2}s_{ab}\,
R_{\mu\nu}{}^{ab}\,\dot{z}^{\nu} + \p_a T_{\mu
\nu}{}^a\,\dot{z}^{\nu}. \ee This identity is satisfied for
arbitrary particle  if \be
R_{\mu\nu}{}^{ab}-\frac{1}{l^2}e_{\mu}{}^{a}\,e_\tau{}^b =0, \ \ \
T_{\mu \nu}{}^a=0 \ee i.e.\  if the background spacetime geometry
is de Sitter. Such geometry holds in the limit in which
$\alpha=0$, eq.\ (\ref{2.2}). In this case the path integral
quantization of the particle can be easily evaluated, see Section
\ref{wilsonlines}.

If one now consider the full gravity case where $\alpha\neq 0$, the
bulk action also breaks the de Sitter invariance down to $\SO(3,1)$.
As we will see in the next section, if one take into account the
variation coming from the term proportional to $\alpha$ ( see
eq. \ref{e4} and comments following it) the equation of motion
obtained by making an $\SO(4,1)$ gauge transformation in the full
gravity theory are consistent with the Mathisson-Papapetrou
equations we derived above. Thus, analogously to what is shown in
the context of pure gravity in \cite{FreidelA} we expect a
formulation of gravity coupled to particles where the the full de
Sitter invariance is manifest, but spontaneously broken down to
$SO(3,1)$ by classical solutions.

\section{Equations of motion for gravity -- particle system}

Let us now consider the gravitational equation of motion when the
gravitational field is coupled to a particle carrying Poincar\'e
charge $\J$ (\ref{2a.7}). The  action of this system will be given
by the sum of actions (\ref{actionpt}) and (\ref{2.1}). The
equations of motion resulting from this action are as follows.
\newline

{\bf $\B$ equations}
\begin{equation}\label{e1}
    B^{ab} = \frac{1}{\alpha^2+\beta^2}\, \left[-\frac\alpha2\, \epsilon^{abcd}\, F_{cd} + \beta\, F^{ab} \right]
\end{equation}

\begin{equation}\label{e2}
  B^{a4} \equiv B^a = \frac1\beta\, F^{a4}= \frac{1}{\beta l}\, d_{\omega}e^{a}
\end{equation}

{\bf $\A$ equations}
\begin{equation}\label{e3}
    \left(\, d_\A\,  \B\right)^{IJ} = \frac{1}{2}  J^{IJ}_P(x)
\end{equation}
where we have introduced the  three-form
\begin{equation}\label{e7}
   J^{IJ}_P(x) = \int \epsilon_{\mu\nu\rho\sigma}\, J^{IJ}(\tau)\dot z^\sigma\, \delta^{4}(x-z(\tau))\, dx^\mu\wedge dx^\nu\wedge dx^\rho
\end{equation}
with $\epsilon_{\mu\nu\rho\sigma}$  the Levi-Civita tensor
$\epsilon_{0123}=1$. This form is such that \be \label{e7a} \int
J_{P}^{IJ}(x)\wedge a(x)= \int_{P}d\tau J^{IJ}(\tau)
\dot{z}^{\mu}(\tau)a_{\mu} \ee for any one form $a$.

Finally we have

 {\bf $\h$
equations}, obtained by varying the action with respect to an
$SO(4,1)$ transformation (\ref{2.3}).
\begin{equation}\label{e4}
   \alpha\, \epsilon^{abcd}\, B_{ab} \wedge B_c =  \left(\int d\tau\,(D_\tau J)^d\, \delta^4(x-z(\tau))\right)\,  d^4x
\end{equation}

Let us  pause for a moment to recall that in the case of a
particle in external,  fixed gravitational field  the $\h$
equation of motion (\ref{hproblem}),  for $\h \in \SO(4,1)$ has
led to constraints imposed on
 components of gravitational field strengths. As we will see in a moment
   this problem is absent if the gravitational field is dynamical,  in the case $\alpha\neq0$. Indeed,
consider eq.\ (\ref{e3}). Applying the covariant derivative to both
sides we get
\begin{equation}\label{e5}
   \left(d_\A\,   d_\A\,  \B\right)^{IJ} =  \frac12(d_\A\, \J_P)^{IJ}(x)=  - \frac12\left(\int D_\tau\J(\tau)\, \delta^4(x-z(\tau))\,  d\tau\right)^{IJ}
\end{equation}
Therefore the component $a4$ of  (\ref{e5}) is just $-1/2$ times the
RHS of eq.\ (\ref{e4}).  Now
$$
\left(d_\A\,   d_\A\,  \B\right)^{IJ} = F^I{}_K \wedge B^{KJ} +
F^J{}_K \wedge B^{IK}
$$
To compare this with eq.\ (\ref{e4}) we just need the translational
component of this:
$$
F^d{}_c \wedge B^{c4} + F^4{}_c \wedge B^{dc} = \left(
B^{cd}-\frac1\beta\, F^{cd}\right)\wedge\frac1{ l} d_\omega\, e_c =
$$
\begin{equation}\label{e6}
 = -\frac{\alpha}{\beta}\frac{1}{\alpha^2 +\beta^2}\left(\alpha \, F^{cd} + \frac\beta 2 \, {\epsilon^{cdab}}\, F_{ab}\right)\wedge \frac{1}{ l}d_\omega\, e_c
\end{equation}
$$
=-\frac{\alpha}{2} \epsilon^{abcd} (B_{ab}\wedge B_{c}).$$ This is
just $-1/2$ times the LHS of eq.\ (\ref{e4}). Thus we conclude that
the $\h$ equation (\ref{e4}) is  just a part of integrability
conditions of  (\ref{e3}). Thus in what follows we can disregard
eq.\ (\ref{e4}) whatsoever.

 We see therefore that in the case, in which gravity is fully
dynamical, it is consistent to take the gauge degrees of freedom
that become dynamical at the particle world-line, described by $\h
\in \SO(4,1)$.
Diffeomorphisms and those $\h$ that belong to $\SO(3,1)$ leave the bulk action invariant,
and therefore they are dynamical degrees of freedom only along the worldline location. The analysis of their dynamics
is not modified by the coupling to gravity and the results of the previous section therefore apply in the case
where gravity is dynamical.
Moreover, these dynamical degree of freedom ensures that the all formalism is invariant under the
usual gauge group of gravity.

Those $\h$ that belong to $\SO(4,1)/\SO(3,1)$ do
not leave the bulk action invariant, and therefore they are
dynamical degrees of freedom even in the absence of the particle.
Since their equation of motion is a subset of Einstein equation, this suggests that they are
determined on-shell by the gravitational and particle degree of freedom.
However, this determination which can be performed in the pure gravitational case \cite{FreidelA} is
more involved and not yet known when matter is present.
\newline

Let us now consider the equation for gravitational field produced by
a point particle in full generality. Our starting point will be eq.\
(\ref{e1}), (\ref{e2}), (\ref{e3}), Consider first  equation
(\ref{e3}) in `rotational' direction and  expand its RHS. \bes
\left(d_\A\, \B\right)^{ab} &=& d_{\omega}B^{ab}+ A^{a4}\wedge B_{4}{}^{b} +A^{b4}\wedge B^{a}{}_{4}\nonumber \\
&=& d_{\omega}B^{ab} +\frac{1}{\beta  l^{2}}\left(T^{a}\wedge e^{b}-
e^{a}\wedge T^{b}\right). \ees Now, from definition of $\F$
(\ref{fdecomp})
\begin{equation}\label{bianchi1}
d_\omega\, F^{ab} = d_\omega\, R^{ab} - \frac{1}{ l^2}\left(T^a
\wedge e^b - T^b \wedge e^a\right) =  - \frac{1}{ l^2}\left(T^a
\wedge e^b - T^b \wedge e^a\right)
\end{equation}
Thus
$$
\left(d_\A\, \B\right)^{ab}= d_{\omega}(B^{ab}
-\frac{1}{\beta}F^{ab}) =-\frac{\alpha}{\beta}\frac{1}{\alpha^2
+\beta^2} d_{\omega}\left(\alpha \, F^{ab} + \frac\beta 2 \,
{\epsilon^{abcd}} F_{cd}\right) =\frac{1}{2}J_{P}^{ab}
$$
and  eq.\ (\ref{e3}) can be written as \be T^a \wedge e^b - T^b
\wedge e^a = \frac{\beta  l^2}{2 \alpha}(\alpha
\delta^{ab}_{cd}-\frac{\beta}{2}\epsilon^{ab}{}_{cd}  ) J^{cd}_{P}
\ee This is an algebraic equation which fully determine the torsion
in term of the  spin of the particle. When written in terms of the
gravitational  parameter it reads \be T^a \wedge e^b - T^b \wedge
e^a = \frac{G \gamma}{2(1+\gamma^2)}(
\delta-\frac{\gamma}{2}\epsilon)^{ab}{}_{cd}\, s^{cd}_{P}
\label{eqg1}\ee and we see that the Immirzi parameter affects the
coupling between torsion and spin; in the case of usual metric
gravity $\gamma=0$ the torsion is zero.

If we now consider the translation part of (\ref{e3}) we get \bes
\left(d_\A\, \B\right)^{a4}&=& d_{\omega}B^{a4}+ A^4{}_{b}\wedge B^{ab}\nonumber\\
&=& \frac{1}{ l}\left(\frac{1}{\beta}d_{\omega}T^{a}-B^{ab}\wedge e_{b}\right)\nonumber\\
&=& \frac{1}{ l}\left(\frac{1}{\beta}F^{ab}-B^{ab}\right)\wedge
e_{b} \ees Thus the field equation gives us the Einstein equation
\be \frac{\alpha}{(\alpha^2+\beta^2)}\left( \frac{\alpha}{\beta}\,
R^{ab} \wedge e_b + \, G^a\right) = \frac{ l}2 J^a_{P} \label{Ee}\ee
where
$$
G^a \equiv \frac{1}{2}\epsilon^{abcd}\, F_{cd}\wedge  e_b  =
\frac{1}{2}\epsilon^{abcd}\,  (R_{cd}\wedge
e_{b}-\frac{\Lambda}{3}e_{b}\wedge e_{c}\wedge e_{d})
$$
is the Einstein tensor with cosmological constant. The first term on
the RHS is the derivative of the torsion $d_{\omega}^2\,
e^{a}=d_{\omega}T^{a}=R^{ab} \wedge e_b $. Equation (\ref{Ee})
written in terms of the gravity coupling constant (\ref{parameter})
reads \be
 \frac{1}{\gamma}\, R^{ab} \wedge e_b + \, G^a = \frac{G}{2} p^{a}_{P}\label{eqg2}
\ee Equations (\ref{eqg1}), (\ref{eqg2}) characterize the
gravitational field produced by point particle with momentum $p^a$
and spin $s^{ab}$, and have the standard form.

\section{Quantum Particles and Wilson lines}\label{wilsonlines}

Let us now describe the effect of quantizing  particle degrees of
freedom. We first restrict our analysis to case of the pure
BF-theory ($\alpha=\beta=0$) to show that in the quantum regime
inclusion of the particle is realized exactly by insertion of Wilson
lines. Indeed in this case the bulk action is invariant under
$\SO(4,1)$ gauge transformation, while the insertion of the particle
breaks this symmetry at the location of the particle. The $\SO(4,1)$
gauge group element $\h(\tau)$ at the location of the particle
becomes a dynamical object and its integration  in the path integral
gives
 \be\label{Path}
 W(A) =\int {\cal D}(\h) e^{i \int_P -\tr \left(\K
\A^\h_\tau(\tau)\right) } \ee with $\A^\h=\h^{-1} \A \h + \h^{-1} d
\h$. We can interpret this path integral as the quantum amplitude
associated with the system described by the particle action
\begin{equation}
S(\h) =  \int {\mathrm d}\tau \tr(\h^{-1}\dot \h \K) + \tr(\A \J ),
\label{actpart}
\end{equation}
where $J=\h K \h^{-1}$.

The phase space variables of this system are $(\h, \J)$ satisfying
the commutation relations (see \cite{desousa}) \be \{\h,J_{IJ}\} =
\gamma_{IJ} \h, \quad \{\J_{IJ},\J_{KL}\} = \eta_{JK} \J_{IL}
+\cdots \ee and subject to the constraints \be C\equiv \K - \h
J\h^{-1} = 0. \ee Among these constraints, 2 are first class
$(C_{04},C_{23})$ and the others are second class. The algebra of
gauge invariant observables commuting with the constraints is
generated by $J_{IJ}$ and their Dirac bracket is equal to the
original bracket. They form an $\SO(4,1)$ algebra subject to the
first class constraints (\ref{C1}), (\ref{C2}) which label a
representation of mass $m$ and spin $s$. Due to the Feynman-Kac
correspondence the path integral (\ref{Path}) computes, for a closed
loop, the trace in the $\so(4,1)$ representation $(m,s)$ of the
Wilson line operator \be
 W(A) = \tr_{(m,s)}(P \exp \int_P A_{IJ}J^{IJ}).
\ee
This correspondence between path integral and Wilson lines has been studied
 in the compact group case by Alekseev et al \cite{Alek}.

Let us remark that this gives an interesting perspective on Feynman
amplitudes: Such amplitudes are related to  path integrals
(\ref{Path}) along the Feynman graph. The Feynman graph amplitude
can thus be interpreted as a spin network evaluation (which
generalizes Wilson loops) of a $\SO(4,1)$ spin network whose edges
are labeled by pairs $(m,s)$.

Now if we turn on gravity $\alpha\neq 0$ the quantization story is  more
involved. In this case the bulk action is invariant only under
$\SO(3,1)$ gauge transformations and the `Wilson line' is now
deformed by the coupling to the gravitational field, it depends on
the value of the $B$ fields and becomes \be\label{Pathg}
 W(A) =\int {\cal D}(\h) e^{i \int_P -\tr \left(\K
\A^\h_\tau(\tau)\right) + \alpha \int_M \tr\left(\gamma(\h)B\wedge B
\right)} \ee where $\gamma(\h)= \h\gamma \h^{-1}$.

\section{Conclusions}

In this paper we investigated the coupling of point particle to
gravity regarded as a constrained topological field theory. Our
results presented here could be treated as a starting point for
various directions of investigations.

First, since the $\alpha$ parameter is small, we can consider a
perturbation theory of gravity coupled to particle(s) being the
perturbation theory in $\alpha$. The distinguished feature of this
theory would be that it is, contrary to earlier approaches,
manifestly diffeomorphism-invariant, so its framework it is possible
to talk about weak gravitational field in the conceptual framework
of full general relativity. These investigations, both in the case
of $\beta=0$ and $\beta\neq0$ will be presented in the forthcoming
paper.

The fuller control over the small $\alpha$ sector will presumably
make it possible to address the outstanding question of what is the
flat space limit of the theory of quantum gravity, coupled to point
particles. It has been claimed that such a theory will be not the
special relativity, but some form of doubly special relativity (see
e.g.\ \cite{Kowalski-Glikman:2006vx}.). This has been shown in the
case of three dimensional gravity \cite{FreidelL}.
Indeed the key ingredient that allowed to construction the effective dynamic of matter coupled to
3d quantum gravity was the understanding that particles can be described as charged topological defect.
This allowed to quantize in one stroke gravity and matter and simplify drastically the problem of
understanding the modification of matter dynamics due to quantum gravity effects.
We hope that the results presented here will similarly simplify the study of quantum gravity to matter fields,
especially in the limit where $\alpha$ is small, and according to the perturbation proposed in \cite{FreidelA}.

Last but not least there is a curious appearance of gravitational
analogues of magnetic monopole in the topological sector of our
theory, with $\alpha=0$, $\beta\neq0$. Consider eq.\ (\ref{e3})
\begin{equation}\label{e8}
    \left(\, d_\A\,  \B\right)^{IJ} = \frac{1}{2}  J^{IJ}_P(x)
\end{equation}
where $J^{IJ}_P(x)$ is defined by (\ref{e7}). Since for $\alpha=0$,
$B^{IJ} = 1/\beta\, F^{IJ}$, it seems for the first sight that eq.\
(\ref{e8})  is simply inconsistent because after substituting $\F$
for $\B$ the left hand side is identically zero by Bianchi identity.
 However having point-like, distributional sources we can allow for
 distributional connections,  for which Bianchi identity does not hold, as in the case of Dirac monopole.

 By construction (cf.\ eq.\
{\ref{J}) there  is always a gauge transformation  which makes it
possible to fix the gauge such that
\begin{equation}\label{e9}
   \frac1\beta\, \left(d_\A \F\right)^{IJ} = \frac{1}{2} K^{IJ}\, \delta_P
\end{equation}
where $\K$ is given by eq.\ (\ref{2a.2}).  It is now clear that a
solution of (\ref{e8}) has the form
\begin{equation}\label{e10}
    \A = \frac{\beta}{2}\, \K\, A_D
\end{equation}
where $A_D$ is the abelian Dirac  monopole connection (see for
example \cite{nakahara}). For example in the patch which does not
cover the negative $z$ axis
$$
A_D = \frac{x\,dy-y\, dx}{r(r+z)}
$$
in cartesian coordinates.

 To construct a general solution let us assume that
\begin{equation}\label{e11}
    \A = \A_0 + \frac{\beta}{2}\, \K\, A_D \equiv \A_0 + \frac{\beta}{2}\, \A_D
\end{equation}
where $\A_0$ is an arbitrary non-singular connection (i.e.,  the one
that satisfies Bianchi identity $d_{\A_0}\, \F(\A_0) = 0$.) Plugging
this anzatz into eq.\ (\ref{e9}) we find that this equation is
satisfied identically.

What is even more interesting, the monopole  configurations
arising in the topological limit of gravity give rise to correct
particle action. To see this, consider the action (\ref{2.1a}) in
the $\alpha=0$ case
\begin{equation}
S=\int \left(B_{IJ}\wedge F^{IJ} -  \frac{\beta}{2}\, B^{IJ}\wedge
B_{IJ}\right) \label{e13}
\end{equation}
Solving for $\B$, and plugging the solution back to the action, we
find
\begin{equation}
S=\frac{1}{2\beta}\,\int \left(F_{IJ}\wedge F^{IJ} \right)
\label{e14}
\end{equation}
Using (\ref{e8}), integrating the delta, and going to arbitrary
gauge we get
\begin{equation}
S=\frac{1}{4}\,\int Tr \left(\A^{\h}(\tau)\, K\right) + CS(\A)
 \label{e15}
\end{equation}
where $CS(\A)$ is the boundary Chern--Simon action on $S^2 \times R$
(spacelike infinity times time.) We see therefore that the particle
action arises from singularities of  topological remnant of
gravitational field, of the form of generalized monopoles. It is not
clear however, if this type of construction can be extended to the
full theory, with $\alpha\neq0$.

\section*{Acknowledgements}

For JK-G this work  is partially supported by the grants KBN 1 P03B
01828 and University of Wroclaw 2594/W/IFT.

\section*{Appendix}

In this appendix, we recall our conventions and present some useful formulas.

For gamma-matrices, we use
\begin{equation}\label{a1}
    \{\gamma^a, \gamma^b\} = 2\, \eta^{ab}
\end{equation}
where $\eta^{ab}$ is the Minkowski metric of signature $(-,+,+,+)$.
For commutator of two gamma matrices, we use the notation
\begin{equation}\label{a2}
  \gamma^{ab} = \frac12\,  [\gamma^a, \gamma^b]
\end{equation}
We denote ``gamma-five'' matrix by $\gamma$:
\begin{equation}\label{a3}
    \gamma = -i \gamma^0\gamma^1\gamma^2\gamma^3, \quad \{\gamma, \gamma^a\}=\{\gamma, \gamma^{ab}\}=0,
\end{equation}
which satisfies $\gamma^2=1$.
The matrices $$\gamma^{a_1\ldots a_n} =
\frac1{n!}\, \left(\gamma^{a_1} \cdots \gamma^{a_n} \pm \mbox{permutations}\right)$$ satisfy the identities
\begin{equation}\label{a4}
   \gamma^{a} \gamma^{b} = \gamma^{ab}+\eta^{ab}
\end{equation}
\begin{equation}\label{a5}
    \gamma^{ab}\gamma^{c} = \gamma^{abc}+\eta^{bc}\gamma^{a}-\eta^{ac}\gamma^{b}
\end{equation}
\begin{equation}\label{a6}
    \gamma^{abc}\gamma^{d} = \epsilon^{abcd}\gamma+\eta^{dc}\gamma^{ab}-\eta^{bd}\gamma^{ac}-\eta^{ad}\gamma^{cb}
\end{equation}
\be
\gamma_{ab} \gamma =\f{i}{2} \epsilon_{abcd}\gamma^c \gamma^d, \, \gamma_{abc} = i \epsilon_{abcd}\gamma^d \gamma
\end{equation}
\be
\{\gamma_a\gamma, \gamma\}=0,\, \{\gamma_{ab},\gamma\}= i\epsilon_{abcd}\gamma^{cd}
\ee
\be
\{\gamma_b\gamma, \gamma_a\}=i\epsilon_{abcd}\gamma^{cd},\,
\{\gamma_{bc},\gamma_a\}= 2i\epsilon_{abcd}\gamma^{d}\gamma
\ee

\begin{equation}\label{a7}
    \gamma^{a}\gamma_{a}=4, \quad \gamma^{a_1\ldots a_n b}\gamma_{b} =(4-n)\,
    \gamma^{a_1\ldots a_n }\quad \gamma^{b}\gamma^{a_1\ldots a_n }\gamma_{b} = (-1)^n\,
    (4-2n) \gamma^{a_1\ldots a_n }
\end{equation}

The Lie algebra $\so(4,1)$ is generated by
$T_{ab} = \gamma_{ab}/4$ and $T_a=\gamma_a\gamma/4$ which satisfy
\bes
2 {[}T_{ab},T_{cd}{]}&=& \eta_{bc} T_{ad} -\eta_{ac} T_{bd}-\eta_{bd} T_{ac}+\eta_{ad} T_{bc},\\
 2 {[}T_{ab},T_{c}{]}&=& \eta_{bc} T_{a} -\eta_{ac} T_{b},\\
  2 {[}T_{a},T_{b}{]}&=& - T_{ab}.
 \ees
Moreover
\be
\tr(T_{ab}T_{cd})=\frac14(\eta_{bc}\eta_{ad}-\eta_{ac}\eta_{bd}),\quad \tr(T_aT_b)=-\frac14\eta_{ab}.
\ee
The normalisation of the generators is such that if $\A=A^{IJ}T_{IJ}$, $\B=B^{IJ}T_{IJ}$ then
\be
 {[}\A,\B{]}= (A^{IK}B_{K}{}^J -B^{IK}A_{K}{}^J )T_{IJ}= {[}\A,\B{]}^{IJ}T_{IJ}.
 \ee

\end{document}